\documentclass[10pt]{iopart}     

\begin{document}

\title[The Hartle-Thorne circular geodesics]{Circular geodesics in the Hartle-Thorne metric} 
\author{M. A. Abramowicz * \P\ \S\,   G.J.E. Almergren * \dag\,
        W.~Klu{\'z}niak *  \ddag\, A.V. Thampan \dag\ }

\address{\P\ Department of Astronomy and Astrophysics, Chalmers University, \\ \quad S-412 96 G\"oteborg, Sweden}
\address{\S\ Silesian University at Opava, Department of Physiscs, \\ \quad CZ-74601 Opava, Czech Republic}
\address{\dag\ International School for Advanced Studies (ISAS/SISSA), \\ \quad Via Beirut 2-4, I-34013 Trieste, Italy}
\address{\ddag\ Institute of Astronomy, University of Zielona G\'ora, \\ \quad Lubuska~2, PL-65-265 Zielona G\'ora, Poland}
\address{\ddag Copernicus Astronomical Center,
 ul. Bartycka 18, 00-716 Warszawa, Poland}
\address{* UK Astrophysical Fluids Facility (UKAFF), Leicester University, England}


\begin{abstract}  
The Hartle-Thorne metric is an exact solution of vacuum Einstein field
equations that describes the exterior of any slowly and rigidly rotating,
stationary and axially symmetric body. The metric is given with accuracy
up to the second order terms in the body's angular momentum, and first order
in its quadrupole moment. We give, with the same accuracy, analytic
formulae for circular geodesics in the Hartle-Thorne metrics. They
describe angular velocity, angular momentum, energy, epicyclic
frequencies, shear, vorticity and Fermi-Walker precession. These
quantities are relevant to several astrophysical phenomena, in particular
to the observed high frequency, kilohertz Quasi Periodic Oscillations (kHz
QPOs) in the X-ray luminosity from black hole and neutron star sources. It
is believed that kHz QPO data may be used to test the strong field regime of
Einstein's general relativity, and the physics of super-dense matter of which
neutron stars are made of. 
\end{abstract} 

\pacs{???}   
\submitto{\CQG} 

\ead{joal@sissa.it}
\ead{marek@fy.chalmers.se}
\ead{wlodek@camk.edu.pl}

\maketitle

\newcommand*{\dd}{{\textrm d}}                       
\newcommand*{\g}{{\mbox{\textsl g}}}                 
\newcommand{\oo}[1]{{\textstyle{\frac{1}{#1}} }}     
\newcommand*{\pd}{\partial}                          
\newcommand*{\pD}[2]{\frac{\pd{#1}}{\pd{#2}}}        
\newcommand*{\yR}{{\mathcal R}}

\newcommand*{\lab}{\label}
\newcommand*{\hl}{\hrulefill \noindent \\}
\newcommand*{\once}{\vspace{.20cm}}
\newcommand{\non}{\nonumber}

\newcommand*{\D}{\displaystyle}
\newcommand*{\T}{\textstyle}
\newcommand*{\Sc}{\scriptstyle}
\newcommand*{\Ss}{\scriptscriptstyle}

\newcommand{\beq}{\begin{equation}}
\newcommand{\eeq}{\end{equation}}
\newcommand{\1}{\beq}
\newcommand{\2}{\eeq}
\newcommand{\3}{\begin{eqnarray}}
\newcommand{\4}{\end{eqnarray}}
\newcommand{\5}{\begin{eqnarray*}}
\newcommand{\6}{\end{eqnarray*}}
\newcommand{\7}[1]{\begin{array}{{#1}}}
\newcommand{\8}{\end{array}}

\newcommand*{\guv}{g_{ab}}
\newcommand*{\gUV}{g^{ab}}
\newcommand*{\Tuv}{T_{ab}}

\setlength{\mathindent}{0.0cm}

\section{The astrophysical motivation}

Thanks to precise timing of binary pulsars' orbital decay 
\cite{TFW}, the weak field
limit of Albert Einstein's general relativity is far better tested than
any other physical theory. However, the most interesting and bizarre
predictions of Einstein's theory deal not with the weak field, but with
the extremely strong field regime and these have never been tested. Thus,
the very question \cite{Will} ``Was Einstein right?''
 remains unanswered. One may argue
that in the foreseeable future there will be no way to test super-strong
gravity. Obviously, this cannot be done in laboratories on Earth. Central
regions of Galactic sources containing black holes and neutron star do
have gravity sufficiently strong for such tests, but they are only several
tens of kilometer across and typically observed from kiloparsecs away.
Thus, they cannot be spatially resolved with current instruments. However,
already existing technology \cite{KLS} provides very precise resolution 
in time of the observed variability of X-ray radiation coming from the
vicinity of black holes and neutron stars.

Of special importance here are the kilohertz, Quasi Periodic Oscillations
(kHz QPOs) of the X-ray flux from galactic low mass X-ray binaries in
which the compact object is either a black hole or a neutron star
\cite{Klis}. 
The QPOs are often observed in twin pairs of kilohertz frequencies. 
The observed frequency ratios are rational, with the particular ratio 3:2
being most common. In several sources it was possible to detect the
orbital binary motion, and deduce the mass of the compact object from
Kepler's laws. In a few cases the deduced mass is above the Chandrasekhar
limit, suggesting a black hole.

Rational ratios of oscillation frequencies suggest a resonance. There is a
strong reason to think of a particular resonance here, the non-linear
parametric resonance between the vertical and radial epicyclic
oscillations in accretion disks 
\cite{MA5}. The world lines of matter in accretion
disks differ only slightly from those corresponding to circular geodesics.
The 3:2 resonance occurs at a precise location inside the disk, only a few
gravitational radii from the source, i.e., in the super-strong gravity
regime. This gives a unique opportunity for observational strong-gravity
tests. In particular, for three black hole twin kHz QPO sources with
known masses, the resonance model has already been used for direct
measurements of the black hole spin 
\cite{MA3,MA4}. That was possible due to the fact
that the black hole metric is known analytically (the Kerr solution).

For neutron and strange stars, the exterior metric is not analytically
known in general. It is known only in the special but important case of a
perfect fluid, stationary and axially symmetric star with mass $M_0$,
angular momentum $J_0$, and quadrupole moment $Q_0$, that rotates rigidly and
slowly. In this case, the Hartle \& Thorne \cite{HT2} solution gives the
metric analytically, with accuracy up to the second order terms in stellar
dimensionless angular momentum $j = J/M^2$, and first order in terms of
dimensionless quadrupole, $q =- Q/M^3$. 

We give here, with the same accuracy, analytic formulae for circular
geodesics in the Hartle-Thorne metrics. They describe angular velocity,
angular momentum, energy, epicyclic frequencies, shear, vorticity and
Fermi-Walker precession. These quantities are of interest in modeling
of several astrophysical phenomena, in particular of the kHz QPOs 
(e.g., \cite{AAKT}). The radius
of the innermost stable marginally orbit and related quantities
have previously been computed to linear order in $j$ 
by Klu\'zniak \& Wagoner~\cite{KW},
through quadratic and higher orders in $j$ and $q$ 
by Shibata \& Sasaki~\cite{ShS}, Sibgatullin \& Sunayev~\cite{SS},
and numerically by Berti \& Stergioulas~\cite{BS}.

Although the derivation of the following formulae is simple, it is technically 
troublesome and one must worry about typographical errors. We derived all
the formulae in a semi automatic way \cite{joal} using Mathematica. 

\section{The Hartle-Thorne Metric}

We use the geometrical units, in which $c = 1 = G$. 
In these units, mass, length and time are all measured in centimeters. 
Adding an asterisk ($^*$) as the indicator of the usual physical units, we write for 
the mass, radius and time in geometrical units, $M = GM^*/c^2$, $r = r^*$, 
and $t = ct^*$. In these units and with spherical coordinates 
$(t, r, \theta, \phi)$, the Hartle-Thorne metric \cite{HS,HT1,HT2} reads,

\1 \fl \dd s^2 = \g_{tt}\,\dd t^2 + \g_{rr}\,\dd r^2 + \g_{\theta \theta}\,\dd \theta^2 
             + \g_{\phi\phi}\,\dd \phi^2 + \g_{\phi t}\,\dd \phi\,\dd t
	                                 + \g_{t \phi}\,\dd t\,\dd\phi \,,\2
where $\g_{t \phi}=\g_{\phi t}$ and the components of the metric tensor are given below.
%
\1 \7{lcr} 
\g_{tt}       &=& + \left(1 - 2M/r \right)
                                     \,\left[ 1 + j^2 \,F_1 + q\,F_2 \right]  \\
\g_{rr}       &=& - \left(1 - 2M/r \right)^{-1} 
                                     \,\left[ 1 + j^2 \,G_1 - q\,F_2 \right] \\
\g_{\theta \theta} &=& - r^2\,         \left[ 1 + j^2 \,H_1 + q\,H_2 \right] \\
\g_{\phi\phi} &=& -r^2 \sin^2 \theta \,\left[ 1 + j^2 \,H_1 + q\,H_2 \right] \\
\g_{t\phi}    &=& -2 (M^2/r) \,j\,\sin^2 \theta \\
\8 \label{eq:TheMetric}\2
%
where 
%
\3 F_1 &=&  \left[ 8 M r^4 (r - 2M) \right]^{-1} \,\cdot \non \\
       & &  \left[ u^2 \,\left( 48 M^6 - 8 M^5 r - 24 M^4 r^2 - 30 M^3 r^3 
               - 60 M^2 r^4 + 135 M r^5 - 45 r^6 \right) \right.\non \\
       & &  \left. + (r - M) \left( 16 M^5 + 8 M^4 r - 10 M^2 r^3 - 30 M r^4 + 15 r^5 
                         \right) \right] + A_1(r) \\
   F_2 &=& \left[ 8 M\,r\,\left( r - 2M \right) \right]^{-1} 
           \left( 5\,(3 u^2 - 1)\,(r - M)\,(2 M^2 + 6 M r  - 3 r^2) \right) - A_1(r) \\
   G_1 &=&  \left[ 8 M\,r^4\, \left( r - 2M \right) \right]^{-1} 
            \left( \left( L - 72 M^5 r\right) 
                   - 3 u^2\,\left( L - 56 M^5 r\right) \right) - A_1(r) \\
   L   &=&  \left( 80 M^6 + 8 M^4 r^2 + 10 M^3 r^3 + 20 M^2 r^4 
                          - 45 M r^5 + 15 r^6 \right) \\[1ex]
   A_1 &=&  \frac{15 r\,(r - 2M)\,( 1 - 3 u^2 )}{16 M^2} \,
            \ln\,\left(\frac{r}{r - 2M}\right) \4
\3 H_1 &=& ( 8 M r^4 )^{-1} \,(1 - 3 u^2) \, 
           ( 16 M^5 + 8 M^4 r - 10 M^2 r^3 + 15 M r^4 + 15 r^5 ) + A_2(r)  \\
   H_2 &=& ( 8 M r )^{-1} \,\left( 5\,(1 - 3 u^2)\,(2 M^2 - 3 M r - 3 r^2) 
                            \right) - A_2(r) \\[1ex]
   A_2 &=&  \frac{15\,(r^2 - 2 M^2)\,(3 u^2 - 1)}{16 M^2} \,
            \ln\,\left(\frac{r}{r - 2M}\right) \4
and where in addition $u := \cos \theta$.

We have checked by directly calculating the Ricci tensor $R_{ik}$, that the 
above Hartle-Thorne metric is indeed a solution of the vacuum Einstein's 
field equations: with the relevant accuracy to quadratic terms in $j$ and 
linear terms in $q$, $R_{ik} = 0$.   

The Kerr metric in the Boyer-Lindquist coordinates could be obtained from 
the above Hartle-Thorne metric after putting $a = Mj$, $q = j^2$, and 
making a coordinate transformation implicitly given by,
\3 r_{\Ss BL}      &=& r - a^2/(2 r^3)\,\left( (r + 2 M)\,(r - M) 
                         + u^2 (r - 2 M)(r + 3 M) \right) \label{eq:C2:HTBL}  \\
\theta_{\Ss BL}    &=& \theta - a^2/(2 r^3)\,(r + 2M)\,\cos\theta\,\sin\theta \4

\section{The detailed formulae}

\subsection{The Horizon}
The radius of the horizon (if present) is obtained by setting 
$\g^2_{\phi t} - \g_{tt}\g_{\phi\phi} =0$ and solving for $r$. 
\1 r_{\Ss h} = 2M\,\left[ 1 - \frac{1}{16}\,j^2\,(7 - 15 u^2)
			    + \frac{5}{16}\,q  \,(1 - 3 u^2) \right]\,.\2 
%
\subsection{The Ergosphere}
Similarly, the radius of ergosphere can be easily obtained 
by setting $\g_{tt}=0$,
\1 r_0 = 2M\,\left[ 1 - \frac{1}{16}\,j^2\,(3 - 11\,u^2)
		      + \frac{5}{16}\,q\,  (1 - 3\,u^2) \right]\,\2 
%
\subsection{Dragging of Inertial Frames}
The angular velocity of an inertial observer near the rotating star as
observed by someone at infinity (i.e. the frame-dragging) is given by:
\1 \omega = - \frac{\g_{t\phi}}{\g_{\phi\phi}} 
          = \frac{2J}{r^3} = \frac{2 M^2 j}{r^3} .\2 
As such, it is independent of the quadrupole moment and the second order terms 
as well as the angle of inclination. 

\subsection{The Orbital Angular Velocity ($\Omega = u^\phi/u^t$)}

The angular velocity for corotating/counterrotating circular particle orbits
is given by:
\1 \Omega  = \pm \frac{M^{1/2}}{r^{3/2}}
               \,\left[ 1 \mp j\,\frac{M^{3/2}}{r^{3/2}} 
               + j^2\,F_1(r) + q\,F_2(r) \right] \2
where
\3 F_1(r)    &=& \left[ 48\,M^7 - 80\,M^6\,r + 4\,M^5\,r^2 - 
                 18\,M^4\,r^3 + 40\,M^3\,r^4 + 10\,M^2\,r^5 \right.\\[1ex]
	     & & {} \left. + 15\,M\,r^6 -  15\,r^7 \right] \,  
                 (16\,M^2\,\left( r - 2M \right) \,r^4)^{-1} + A(r) \\[1ex]
   F_2(r)    &=& \frac{5\,\left( 6\,M^4 - 8\,M^3\,r - 2\,M^2\,r^2 - 
                 3\,M\,r^3 + 3\,r^4 \right) }{16\,M^2\,
                  \left( r - 2M \right) \,r} - A(r) \\[1ex]
   A(r)      &=& \frac{15\,\left( r^3 - 2\,M^3\right)}{32\,M^3} \,
                 \ln \left(\frac{r}{r - 2M}\right) \4
%
\subsection{The Specific Angular Momentum ($\ell = - u_\phi/u_t$)}
%
\1 \ell = \pm \ell_0 \left[ 1 
          \mp j\,\frac{M^{3/2}(3r-4M)}{r^{3/2}(r-2M)} 
          + j^2 \,F_1(r) - q\,F_2(r) \right] \2
where
\3 \ell_0 &:=& \frac{ M^{1/2} r^{3/2} }{r - 2 M} \\[1ex]
   F_1(r) &=&  \left[ 16 M^2\,r^4\, \left( r - 2M \right)^2 \right]^{-1} 
               \left( 96 M^8 - 112 M^7 r - 8 M^6 r^2 - 48 M^5 r^3\right.
                    \non \\
          & & \left. + 42 M^4 r^4 + 220 M^3 r^5 - 260 M^2 r^6 
                     + 105 M r^7 - 15 r^8 \right) + A(r)  \\
   F_2(r) &=& \left[ 16 M^2\,r\,\left( r - 2M \right) \right]^{-1}  
	      \,5\,\left( 6 M^4 - 22 M^2r^2 + 15 M r^3 
                     -3 r^4 \right) + A(r) \\
   A(r)   &=&  \frac{15}{32 M^3} \,
               \left( 2 M^3 +4 M^2 r - 4 M r^2 + r^3 \right) \,
               \ln\,\left(\frac{r}{r - 2M}\right) \4
%
\subsection{The Specific Energy ($\varepsilon = u_t$)}
%
\1 \varepsilon = E_0 \left[ 1 \mp j\,F_1(r) + j^2\,F_2(r) +q\,F_3(r) 
                 \right] \2
where
\3 E_0  &:=& \frac{r-2M}{r^{1/2} (r-3M)^{1/2}}  \\[1ex]
   F_1(r) &=& \frac{M^{5/2}}{r^{1/2}\,(r - 2M)(r - 3M)} \\
   F_2(r) &=& \left[ 16 M\,r^4\,(r -  2 M)\,(r - 3M)^2 \right]^{-1} 
              \left( 144 M^8 - 144 M^7 r - 28 M^6 r^2 \right.
	      \\ \nonumber
          & & \left. - 58 M^5 r^3 - 176 M^4 r^4 +  685 M^3 r^5 - 610 M^2 r^6 
                     + 225 M r^7 - 30 r^8 \right) + B(r)  \\
   F_3(r) &=& \frac{ 5\,(r - M)\,( 6 M^3 - 20 M^2 r - 21 M r^2 + 6 r^3) }
                   { 16 M\,r\,(r - 2M)\,(r - 3M) } - B(r) \\
   B(r) &=&  \frac{15 r\,\left( 8 M^2 -7 M r +2 r^2 \right)}{32 M^2\,(r - 3M)}
             \ln\,\left(\frac{r}{r - 2M}\right) \4
%
\subsection{Radius of marginally stable, marginally bound and photon orbit.}
The condition $\varepsilon = u_t = 1$ gives the radius of the marginally 
bound orbit, $r_{mb}$ and the condition $u^t=0$ gives the photon orbit,
$r_{ph}$. In addition by setting $d\ell/dr = 0$ we can solve for the radius 
of the marginally stable orbit, $r_{ms}$.

\3 \fl r_{mb} &=& 4M \, \left[ 1 \mp \oo{2}\,j 
                - j^2 \, \left( 8047/256 - 45 \ln 2 \right)
	        + q   \, \left( 1005/32  - 45 \ln 2 \right) \right] \\[1ex] 
   \fl r_{ph} &=& 3M \,\left[ 1 \mp j\,\frac{2\,\sqrt{3} }{9} 
                - j^2 \,\left( \frac{7036 - 6075 \ln 3 }{1296} \right) 
	        + q   \,\left( \frac{7020 - 6075 \ln 3 }{1296} \right) \right] \\[1ex] 
   \fl r_{ms} &=& 6\,M\,\left[ 1 \mp j \,\frac{2}{3} \sqrt{\frac{2}{3}} + 
                  j^2\,\left( \frac{251647}{2592} 
                 - 240\,\ln \,\frac{3}{2} \right)
                 + q\,\left( -\,\frac{9325}{96}  
                 + 240\,\ln \,\frac{3}{2} \right) \right] \4
%

\subsection{The Epicyclic Frequencies}
We define an effective potential as:
$U(r,\theta,\ell) := \g^{tt} - 2\ell\,\g^{t\phi} + \ell^2\,\g^{\phi\phi}$,
which can be used to find the general formula for the epicyclic frequencies
on the equatorial plane.
\1 \widetilde{\kappa}^2_x =
   \frac{(\g_{tt} +\Omega_\pm\,\g_{t\phi})^2}{2\g_{xx}}\,
   \left(\pD{^2U}{x^2}\right)_{\ell},
    \qquad \qquad \mbox{$x\ \epsilon\ (r,\theta)$} \2 
Then we have:
\3 \widetilde{\kappa}_r^2      &=& M (r - 6M) r^{-4} \, 
          \left[ 1 \pm j\, F_1(r) - j^2\, F_2(r) - q\, F_3(r)\right] \\
   \widetilde{\kappa}_\theta^2 &=& M r^{-3} \, 
          \left[ 1 \mp j\, G_1(r) + j^2\, G_2(r) + q\, G_3(r) \right] \4
where
\3 F_1(r) &=& \frac{6\,M^{3/2}\,(r + 2M)}{r^{3/2}\,(r - 6M)} \\
   F_2(r) &=& \left[8 M^2\,r^4\,(r - 2M)\,(r - 6M) \right]^{-1} 
              \left[384 M^8 -720 M^7 r -112 M^6 r^2 -76 M^5 r^3\right.\non \\
          & & \left. {} - 138 M^4 r^4 - 130 M^3 r^5 + 635 M^2 r^6 
                     - 375 M r^7 + 60 r^8 \right] + A(r)  \\[1ex]
   F_3(r) &=& \frac{5\,( 48 M^5 + 30 M^4 r + 26 M^3 r^2 - 127 M^2 r^3 
               + 75 M r^4 - 12 r^5)}{8 M^2\,r\,(r - 2M)\,(r - 6M)} -A(r)\\
   A(r)   &=& \frac{15 r\,(r - 2M)\,(2 M^2 + 13 M r - 4 r^2)}{16 M^3\,(r - 6M)}
              \ln\,\left(\frac{r}{r - 2M}\right) \\[3ex]
   G_1(r) &=& \frac{6\,M^{3/2}}{r^{3/2}} \\
   G_2(r) &=& \left[8 M^2\,r^4\,(r - 2M) \right]^{-1} 
              \left[48 M^7 - 224 M^6 r + 28 M^5 r^2 \right. \nonumber \\
          & & \left. {} + 6 M^4 r^3 - 170 M^3 r^4 + 295 M^2 r^5 
                     - 165 M r^6 + 30 r^7 \right] - B(r)  \\[1ex]
   G_3(r) &=& \frac{5\,\left( 6 M^4 + 34 M^3 r - 59 M^2 r^2 + 33 M r^3 
                     - 6 r^4 \right)}{8 M^2\,r\,(r - 2M)} + B(r)  \\
   B(r)   &=& \frac{15 \,(2r - M)\,(r - 2M)^2}{16 M^3}
              \ln\,\left(\frac{r}{r - 2M}\right) \4
%

\subsection{Shear and Vorticity}
The general formulae for shear and vorticity are \cite{MA1}:
\3 \sigma^2 &=& -\oo{4}\,(1 - \Omega \ell)^{-2}\,\yR^2
                 \,(\nabla_a \Omega)(\nabla^a \Omega) \\
   \omega^2 &=& -\oo{4}\,(1 - \Omega \ell)^{-2}\,\yR^{-2}
                 \,(\nabla_a \ell)(\nabla^a \ell) \\[1ex]
      \yR^2 &:=& \left(\frac{U}{E}\right)^2 
                 \,(g_{t\phi}^2 - g_{tt} g_{\phi\phi}) 
              =  \frac{(\ell g_{t\phi} + g_{\phi\phi})^2}{g_{t\phi}^2 
                                           - g_{tt}\,g_{\phi\phi}} \4
The results for the HT metric are:
\3 \sigma^2 &=& S_0 \,\left[ 1 \pm j\,F_1(r) + j^2\,F_2(r) 
                                     + q\,F_3(r)\right] \\
   \omega^2 &=& V_0 \,\left[ 1 \mp j\,G_1(r) + j^2\,G_2(r) 
                                     + q\,G_3(r)\right] \4
where 
\3 S_0  &:=& \frac{9M}{16 r^3}\,\frac{(r - 2 M)^2}{(r - 3 M)^2} \\
   V_0  &:=& \frac{M}{16 r^3} \,\frac{(r - 6 M)^2}{(r - 3 M)^2} \4
and 
\3 F_1(r) &=& \frac{4 M^{3/2}}{r^{1/2} (r-3M)} \\
   F_2(r) &=& \left[ 8 M^2 r^4 \,(r - 2 M)^2 (r - 3 M)^2 \right]^{-1} \cdot \non \\
           && -\left(864 M^{10} - 1776 M^9 r + 1048 M^8 r^2 - 592 M^7 r^3 - 10 M^6 r^4 
                                + 542 M^5 r^5 \right. \non \\
           && \left. {} - 693 M^4 r^6 + 820 M^3 r^7 - 530 M^2 r^8 + 150 Mr^9 - 15 r^{10} 
              \right) - A(r) \\
   F_3(r) &=& \frac{5 (36 M^6 - 2 M^5 r - 28 M^4 r^2 + 35 M^3 r^3 
		 - 43 M^2 r^4 + 21 M r^5 - 3 r^6)}{8 M^2 r 
                 \,(r - 2 M)^2 (r - 3 M)} + A(r) \\[1ex]
   A(r)   &=& \frac{15 r\,(r - 4 M) (M^2 + r^2)}{16 M^3\,(r - 3 M)}
              \,\ln\,\left(\frac{r}{r - 2M}\right) \4

\3 G_1(r) &=& \frac{12 M^{3/2}\,(r - 2M)}{r^{1/2}\,(r - 3M)(r - 6M)} \\
   G_2(r) &=& -\left( 7776 M^{10} - 12528 M^9 r + 3672 M^8 r^2 - 4080 M^7 r^3 
                                  + 150 M^6 r^4 \right. \non \\
           && \left. {} - 12594 M^5 r^5 + 30891 M^4 r^6 - 25620 M^3 r^7 + 9590 M^2 r^8 - 1650 M r^9 
              \right. \non \\
           && \left. {} + 105 r^{10} \right) \cdot \left[8 M^2 r^4\,(r - 3 M)^2\,(r - 6 M)^2\right]^{-1} + B(r) \\[1ex]
   G_3(r) &=& \frac{-5 (54 M^5 + 30 M^4 r + 147 M^3 r^2 - 271 M^2 r^3 
              + 141 M r^4 - 21 r^5 )}{8 M^2 r\,(r - 3 M)\,(r - 6 M)} 
              - B(r) \\[1ex] 
   B(r)   &=& \frac{15\ r\,\left(24 M^4 - 126 M^3 r + 135 M^2 r^2 
              - 54M r^3 + 7 r^4 \right)}{16 M^3\,(r - 3 M)\,(r - 6 M)}
	      \,\ln\,\left(\frac{r}{r - 2M}\right) \4
%
\subsection{The Fermi-Walker Precession}
The precession for a gyroscope on a general circular orbit
in a general stationary and axially symmetric spacetime was calculated
in \cite{MA2}. In the equatorial plane for the Hartle-Thorne metric
we use formula (4.8) of \cite{MA2} to obtain: 
\1 \Pi   = \Pi_0\,\left[-1 \mp j\,F_1(r) + j^2\,F_2(r) + q\,F_3(r)\right] \2
\3 \Pi_0 &:=& \frac{M^{1/2}\,r^2}{(r - 3M)^{1/2}\,(r - 2M)^3} \\[1ex]
F_1(r) &=& \frac{3 M^{5/2}\,(r - 4 M)}{2\,r^{3/2}\,(r - 2M)\,(r - 3M)}\\ 
F_2(r) &=& \left( 576 M^{10} - 1704 M^9 r + 2904 M^8 r^2 
             - 1280 M^7 r^3 + 1604 M^6 r^4 - 8060 M^5 r^5 \right. \non \\
       & & \left. + 10688 M^4 r^6 - 6290 M^3\,r^7 + 1865 M^2 r^8 - 270 M r^9 
             + 15 r^{10} \right)\,\cdot \non \\ 
       & & \left[ 16 M^2\,r^4\,(r - 2M)^2\,(r - 3M)^2 \right]^{-1} 
           - A(r) \\[1ex]
F_3(r) &=& \frac{5\,( 12 M^4 + 36 M^3 r - 94 M^2 r^2 + 33 M r^3 - 3 r^4)
               }{16 M^2\,r\,(r - 3M)} + A(r) \\[1ex]
A(r)   &=& \frac{15 r\,(r - 4M)\,(10 M^2 - 8 Mr + r^2)}{32 M^3\,(r - 3M)}
           \ln \left( \frac{r}{r - 2M} \right) \4
%
\subsection{Resonance}
Assuming an $m:n$ resonance between epicyclic frequencies,
we calculate the radius 
for which the resonance occurs given the condition that 
$m\,\widetilde{\kappa}_r = n\,\widetilde{\kappa}_\theta$ and denoting 
$z := m^2/n^2$,
\1 r_{mn} = r_0 \,\left[1-j\,F_1(m,n)+j^2\,F_2(m,n)+q\,F_3(m,n) \right] \2
where
\1 r_0 = 6\,M\,\left(1-\frac{n^2}{m^2}\right)^{-1} \2
and
\3 F_1 &=& \frac{1}{m^3}\,\left(\frac{2}{3^3}\right)^{1/2}\,(m^2-n^2)^{1/2}\,
           (2 m^2 + n^2) \\
   F_2 &=& \left( 1 - 50\,z - 11\,z^2 + 385\,z^3 + 10612\,z^4 + 123286\,z^5 
           + 496927\,z^6 \right. \non \\
       & & \left. + 691843\,z^7 + 251647\,z^8 \right) \, \cdot 
	   \left[ 1296\,z^4\, \left( z-1\right)^3 \,
           \left( 1 + 2\,z \right) \right]^{-1} - A(z) \\[1ex]
   F_3 &=& \frac{-5\,\left( 1 + 5\,z \right)\,\left( 1 + 74\,z + 546\,z^2 
           + 950\,z^3 + 373\,z^4 \right) }{48\,{\left( z-1\right) }^3\,z\,
           \left( 1 + 2\,z \right) } + A(z) \\
  A(z) &=& \frac{15\,\left( 1 + 2\,z \right) \, \left( 1 + 12\,z + 63\,z^2 
           + 32\,z^3 \right)}{4\,\left( z - 1 \right)^4}\, 
           \ln \left(\frac{3\,z}{1 + 2\,z}\right) \4
%





\section{Software and hardware}
These results were all derived and checked using Mathematica 4.0 \cite{soft:Ma}
running on a PIII $800$ MHz/256MB under Linux Red Hat 7.2.
In addition the packages TTC \cite{soft:TTC} and GRTensorM \cite{soft:GRT} 
were found useful. The Mathematica notebooks containing the calculations can 
be found on:
%
%

\section*{Acknowledgements}
This work was supported by SISSA, Silesian University of Opava, KBN grants
2P03D01424, PBZ-KBN-054/P03/2001, and the European Commission grant
{\it Access to Research Infrastructure of the Improving Human Potential
Program} to UKAFF at Leicester University.

\section*{References}   


\end{document}